\documentclass[reprint,twocolumn,superscriptaddress,floatfix,showpacs,pra]{revtex4-1}

\usepackage{graphicx}
\usepackage{subfigure}
\usepackage{wrapfig}
\usepackage[ps2pdf]{hyperref}
\usepackage{hypernat}

\newcommand{\ket}[1]{\ensuremath{|#1 \rangle}}
\newcommand{\me}{\ensuremath{\mathrm{e}}}
\newcommand{\mi}{\ensuremath{\mathrm{i}}}

\begin{document}

\title{Reduce, reuse, recycle, for robust cluster state generation}
\author{Clare Horsman} 
\altaffiliation[Current address: ]{Keio University Shonan Fujisawa Campus, Kanagawa 252-0882, Japan}
\email{clare@sfc.wide.ad.jp}
\affiliation{Department of Mathematics, University of Bristol, University Walk, Bristol BS8 1TW, UK}
\affiliation{H. H. Wills Physics Laboratory, University of Bristol, Tyndall Avenue, Bristol, BS8 1TL, UK}
\author{Katherine L. Brown}
\affiliation{School of Physics and Astronomy, University of Leeds, Woodhouse Lane, Leeds, LS2 9JT, UK}
\author{William J. Munro}
\affiliation{National Institute of Informatics, 2-1-2 Hitotsubashi, Chiyoda-ku, Tokyo 101-8430, Japan}
\affiliation{NTT Basic Research Laboratories, NTT Corporation, 3-1 Morinosato-Wakamiya, Atsugi-shi, Kanagawa 243-0198, Japan}
\author{Vivien M. Kendon}
\affiliation{School of Physics and Astronomy, University of Leeds, Woodhouse Lane, Leeds, LS2 9JT, UK}
\date{29th March 2011}

\begin{abstract}
Efficient generation of cluster states is crucial for 
engineering large-scale measurement-based quantum computers. Hybrid 
matter-optical systems offer a robust, scalable path to this goal.  Such 
systems have an ancilla which acts as a bus connecting the qubits.  We 
show that by generating smaller cluster ``Lego bricks'', reusing one 
ancilla per brick, the cluster can be produced with maximal efficiency, 
requiring fewer than half the operations compared with no bus 
reuse.  By reducing the time required to prepare sections of the cluster, 
bus reuse more than doubles the size of the computational workspace that
can be used before decoherence effects dominate.
A row of buses in parallel provides fully scalable cluster state
generation requiring only 20 controlled-\textsc{phase} gates per bus use.
\end{abstract}

\maketitle

\section{Introduction}

Hybrid schemes for quantum information processing are among the most 
promising for scalable quantum computers.  Such systems combine both 
matter and optical elements, where the computational gates between qubits 
of one type can be mediated by a shared bus of the other type
\cite{qubus06,loockhybrid,milburn2000}.
A computational model for such hybrid systems has recently
been characterized as \emph{ancilla-based computation} \cite{elham1},
in contrast to the 
usual quantum computation models that use direct qubit-qubit gates. 
Ancilla-driven schemes are important for chip-based quantum 
computing architectures, where a flying ancilla mediates between 
fixed qubits \cite{Devitt2007,raducatalysis,simon2}.

Hybrid architectures 
form a natural substrate for measurement-based quantum computing (MBQC)
\cite{mbqc2}, one type of which (the topological model based on the surface
code \cite{kitaev}) has the best 
error threshold for quantum computing \cite{surfacecode}.
In MBQC
a highly-entangled \emph{cluster state} is generated, and 
then computation performed by sequential qubit measurements. The quantum 
processing task is to generate the cluster state, after which it becomes 
a matter of measurement and classical processing to feed forward the
measurement outcomes.
The first proposal for cluster-state construction was a one
shot scheme, where the entire cluster was created by a small number of
global operations \cite{mbqc2}.
Since the cluster qubits are measured sequentially, in scalable
physical realizations the cluster is prepared 
dynamically, a few rows at a time \cite{devitt10a,stephens08a}.
This avoids the need for long 
coherence times for entangled qubits \cite{nadevreznik,Nielsen2004},
a critical requirement for scalable schemes.
Photonic schemes for constructing cluster states probabilistically
\cite{Nielsen2004} exploit the linear optics quantum computing
scheme of Knill, Laflamme and Milburn \cite{Knill2001}.
The disadvantage of this approach is the large number of repeated operations
required to successfully build the cluster.  To reduce this overhead,
heralded \textsc{cphase} operations occurring between two qubits were proposed
by Browne \textit{et al.}~\cite{Browne2005}.  Duan \textit{et al.}~\cite{Duan2005} showed
that this probabilistic generation does indeed allow the cluster to grow,
and Gross \textit{et al.}~\cite{jens} determined the optimal growth strategies
for regimes with low and high probabilities of success per operation.
Louis \textit{et al.}~\cite{Louis2007} showed that 
using a three qubit entangling gate instead of a two qubit entangling gate
increased the success probability from $1/2$ to $3/4$.
The advantages of deterministic gates were explored by exploiting
ancilla-based schemes
\cite{hutchinson04,Louis2007,Devitt2007,Lin2010,Ionicioiu10a}.  
Wang \textit{et al.}~\cite{Wang2010} proposed a method to transfer an
atomic cluster state to photonic qubits, inverting the usual role of
the qubit and ancilla between the matter and optical systems.

As the cluster state is the fundamental quantum resource of a 
measurement-based computation, it becomes extremely important to make it 
as error-free as possible. Errors in constructing the cluster can 
propagate rapidly through a computation because of the highly-entangled 
nature of the state, leading to failure of the computation.
Topological surface encodings on cluster states provide a robust
fault-tolerance for quantum computation, provided each component
in the system has an error below a certain threshold
\cite{austin1,austin2,simon2}.
The construction of the cluster itself is one such component,
and schemes to reduce cluster error can enable systems that would
otherwise be unusable to reach the threshold for use with error correction.
Hybrid systems are susceptible to specific types of error that other
systems are not, because of the use of the mediating ancilla.
In cases where the ancilla is not destroyed after each gate there
is the additional possibility of errors propagating through ancilla reuse.
We show there is a trade-off between increasing efficiency by using
the same bus for multiple gates, and increasing errors because of this.

In this paper we present the optimal scheme for dynamic 
2D cluster-state generation in hybrid systems where the mediating 
system (bus) can be used for more than one gate operation without being reset.
We divide the cluster state into ``Lego bricks'', each of which is built with a 
single bus. We give the optimal method for constructing the bricks,
reducing the number of system-bus entanglements.
We then show how to determine the brick size based on the 
error threshold of the system being used.
We find that, even when the probability of error in the system is high,
this scheme can still deliver significant efficiency savings through bus
reuse, enabling a larger cluster to be generated.
The paper is organised as follows.  In section \ref{sec:qubus} we
give an overview of the qubus system, the particular ancilla-based
scheme we will focus on.  Section \ref{sec:reuse} explains how to
reduce the number of operations required when reusing the qubus for
multiple gates.  In section \ref{sec:errors} we introduce our error 
model for reusing the qubus, and in section \ref{sec:cluster} we
apply bus reuse to generating a 2D cluster state.  In section \ref{sec:optimal}
we calculate the optimal bus reuse scheme in the ideal
case, and in section \ref{sec:dephase} we combine this with our error model
to give the optimal bus reuse scheme with dephasing.  Section \ref{sec:dynamic}
discusses how to apply our results to dynamic generation schemes,
and in section \ref{sec:threshold} we calculate the optimal brick size
in terms of the system parameters.  Section \ref{sec:conc} summarises
our conclusions.

\section{Qubus system}\label{sec:qubus}

To provide a concrete setting for our calculations, we will
focus on the qubus system, which consists of matter qubits and a photonic field
as the mediating ancilla \cite{qubus3,qubus2,qubus06}.
The cluster state we are generating is a regular square lattice of 
qubits with nearest neighbors entangled. 
Each qubit is initialized in the state $(\ket{0} + \ket{1})/\sqrt{2}$,
then \textsc{cphase} gates are applied between neighboring qubits. 
Using the qubus, \textsc{cphase} gates are performed using a conditional evolution
\begin{equation}
U_e = \exp (-\mi H_{int} \tau/\hbar),
\end{equation}
where $\tau$ is the fixed time for one such 
operation and
\begin{equation}
H_{int} = \hbar \chi \sigma_z (a^\dagger \me^{\mi \theta} 
+ a \me^{-\mi \theta}),
\end{equation}
where $\chi$ is nonlinearity strength, 
$a(a^{\dagger})$ the field annihilation(creation) operators, and $\theta 
= 0(\pi/2)$ describes coupling of the qubit to the position(momentum) 
quadrature of the field.  The result of this interaction is 
deterministic displacements along discrete paths in phase space,
of amplitude  $\beta = \chi \tau$. The application of $U_e(\pm x_j)$ applies
a displacement of $\beta$ in the positive(negative) direction
in position-space for the $j$-th qubit, and $U_e(\pm p_k)$ 
a displacement of $\beta$ along the positive(negative) axis
in momentum-space for the $k$-th qubit. 
The sequence
\begin{equation}
U_e( x_2)U_e(-p_1)U_e(-x_2)U_e(p_1)
\end{equation}
performs a geometric phase gate between qubits 1 and 2,
with the phase change proportional to the area traced out
\cite{qubus06,wangzanardi}.
When $\beta^2 = \pi/8$, this provides the \textsc{cphase} gate required
for cluster-state construction.
The qubus thus acts as a discrete-level system with two partitions, 
equivalent to two coupled qudit ancillas.
There are two options for using such an ancilla-based system
to construct a cluster state.  Either the ancilla is discarded
after every gate, or it is recycled for use with further gates. 

\section{Reusing the bus}\label{sec:reuse}

If each \textsc{cphase} gate is performed by a 
different bus, then each qubit in the cluster (apart from the perimeter) 
needs to be operated on by four different buses to generate the four 
entanglements it is part of.
For a cluster of $m \times n$ qubits we therefore need 
\begin{equation}
N = 8mn - 4(m+n)
\label{eq:Op1}
\end{equation}
bus operations to complete it -- one entangling and one disentangling 
operation per qubit per gate. However, if we are able to reuse the bus, 
then we can use fewer operations. Consider the following sequence of unitaries
for three qubits: 
\begin{equation}
U_e( x_3)U_e(-p_2) U_e(-x_3)U_e(-x_1)U_e( p_2)U_e( x_1).
\label{eq:2CPhase}
\end{equation}
Reading from the right, a \textsc{cphase} is performed between qubits 1 and 2,
and then qubit 1 disentangled from the bus.
Qubit 2 is kept on the bus, and qubit 3 entangled
with the position quadrature. Finally, qubits 2 and 3 are disentangled
from the bus (in that order).  The result is \textsc{cphase} gates between both
$(1,2)$ and $(2,3)$ using six bus operations rather than the 
8 needed if qubit 2 were disentangled after the first interaction. Such 
sequential operations are possible in all ancilla-based systems which 
can reuse the ancilla.

\section{Dephasing errors}\label{sec:errors}

Reusing the ancilla reduces the total number of operations required,
speeding up the process and hence reducing the length of time
decoherence acts on the cluster qubits. 
For $N$ bus operations taking a total time $N\tau$ to perform,
the probability of a phase-flip error due to qubit dephasing 
is $(1-\exp[-N \gamma \tau])/2$, where $\gamma$ is the dephasing rate 
for one qubit.  Fewer bus operations therefore mean less dephasing.
However, we have to 
take into account error accumulating on the ancilla.
For the qubus, the errors come from photon loss.
The probability of a phase-flip error due to photon loss on the 
bus is $(1- \exp[-4 C \eta \beta^2])/2$, where $C$ is the number of 
\textsc{cphase} gates constructed per bus and $\eta$ is the loss parameter for the bus.
Combining these gives the total probability of dephasing:
\begin{equation}
\varepsilon = \frac{1}{2}\left[1 - \exp(-N \gamma \tau-4 C \eta \beta^2)\right].
\label{eq:dephasing}
\end{equation}
We can therefore trade off the two dephasings by reusing 
the bus, which reduces $N$ but increases $C$. If we minimize $N$ for a given $\varepsilon$, this 
then enables a maximum number of \textsc{cphase} gates to be completed 
before the dephasing reaches a critical value.

\section{Cluster-state generation}\label{sec:cluster}

We now apply bus reuse to more efficient cluster-state construction.
Extending the bus reuse sequence in equation (\ref{eq:2CPhase})
to further qubits allows one ancilla to
generate a line of entangled qubits with just two operations 
per qubit -- one entangling followed by one disentangling.
A set of such lines of length $L$ arranged to form a
$L\times L$ grid generates a 2D cluster, as proposed by
Louis \textit{et al.}~\cite{Louis2007,louis2}.
The minimum number of operations required to build a cluster
from 1D entangled lines of qubits can be obtained by a simple
combinatorial argument.  Consider the cluster as a 2D lattice 
graph, with qubits as vertices and entanglement links as edges.
We count how many edges of the graph can be generated using a line of qubits
when each qubit is only visited once: this corresponds to being
connected to the bus once only, thus minimizing bus use.
For the cluster of $m \times n$ qubits, the total number of vertices is $mn$. 
The maximum number of edges that can be generated is therefore $mn-1$. 
Each entangling action requires two bus operations per qubit (one to 
connect to the bus, one to disconnect).  We can therefore generate $mn-1$ 
edges with $2mn$ bus operations.
The total number of edges in the cluster is $m(n-1) + n(m-1)$,
so we are left with $(n-1)(m-1)$ edges to fill in.
The path we have generated can connect a maximum of
two edges to each vertex in the lattice. 
All vertices except the corners require more than two edges.
Therefore all qubits except the four corners will require reactivation
in order to fill in the extra edges, requiring $2mn-8$ additional operations.
We therefore have a minimum number of bus operations $4mn-8$ to 
generate the cluster state using an ancilla system where at most two 
qubits are coupled to the ancilla at any time.
The method of Louis \textit{et al.}~\cite{Louis2007} achieves the minimum
up to a constant.
If the total number of operations one bus can perform is limited
by the errors accumulating, each change to a new bus requires in general
an extra disentangling of the old bus and re-entangling of the new bus, 
a total of two extra operations per extra bus.

\section{Optimal bus reuse}\label{sec:optimal}

Generation of a line of entangled qubits is the simplest use of
sequential operations, with a maximum of one qubit on
each quadrature of the bus at any time.
However, it does not use the full power of the qubus to reduce the
number of bus operations per gate.
The displacement operators on the qubus allow a qubit on one 
quadrature to become entangled with all qubits on the other. If we start 
by connecting one qubit to, say, the position quadrature, then all its 
neighbors can be simultaneously coupled to the momentum quadrature. 
However, if we then try to connect any other qubit to the position 
quadrature there will be cross-entanglements generated that are not part 
of the required cluster state (where qubits are entangled only with 
nearest neighbors). Only two of the momentum quadrature qubits can 
remain on the bus and not generate unwanted entanglement. These qubits 
must neighbor both of the position quadrature qubits; this then forms a 
closed box in the lattice.

\begin{figure}
   \centering
        \includegraphics[width=4cm]{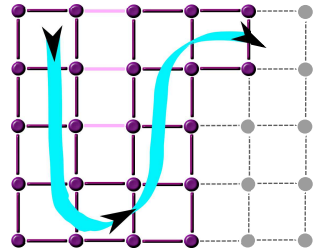}
   \caption{A path of width two generating a $5\times 6$ cluster.
    Dark edges represent entanglement between qubits
    that have connected to the bus once, light edges
    require at least one qubit to connect to the bus twice,
    and dotted edges indicate \textsc{cphase} gates not yet performed
(color online).}
   \label{path2}
\end{figure}

In contrast with the previous scenario, we now need to consider a path 
across the lattice that is two qubits wide, rather than a single qubit 
line (figure \ref{path2}). By inspection, the maximum number of edges
that can be generated on such a path that visits all $mn$ qubits
in the cluster only once is $3mn/2-2$, for even $mn$.
The number of cluster edges remaining after $2mn$ sequential
bus operations is therefore $\frac{1}{2}mn-(m+n)+2$.
We could either finish the sequential operations and then 
generate these edges separately (requiring two qubits per edge to be 
reconnected with the bus), or we could construct these edges as we go 
along.
In the latter case, before a qubit is disconnected from the bus,
we generate the extra edges required for that qubit.
Then only one qubit per additional edge needs to be connected to the bus again.
With two bus operations per connection, this requires
a further $mn - 2(m+n) +4$ uses of the bus.
This gives a lower bound of 
\begin{equation}
N_{\rm{min}} = 3mn - 2(m+n) + 4
\label{eq:Opmin}
\end{equation}
operations to construct the cluster.
We show elsewhere \cite{brown11b} that a using a spiral pattern
for the width two path achieves the bound $N_{\rm{min}}$.
A spiral path does not allow dynamic generation, so
in practice we will use a zig-zag path (figure \ref{path2}).
The U-shaped turns require up to two extra operations per turn,
so we will want to minimise their number to minimise
the actual cost.  The zig-zag path in figure \ref{path2}
is the minimum turn arrangement for dynamically
generating rectangular clusters.

Equation (\ref{eq:Opmin}) tells us the most efficient a scheme can be
when the bus acts as an ancilla partitioned into two.
Clearly in general an ancilla can
have more than two partitions, although multipartition ancillas are
more naturally suited to multiqubit gates. 
With a path of width $a$, the number of operations using a
single bus would only improve to order $2mn + 2(mn/a)$.
We can see then that going to large partition sizes significantly
increases the ancilla complexity for a rapidly reducing 
pay-off in terms of bus efficiency.

\section{Reuse with dephasing}\label{sec:dephase}

\begin{figure}[t]
   \centering
    \subfigure[]
       {\includegraphics[width=4cm]{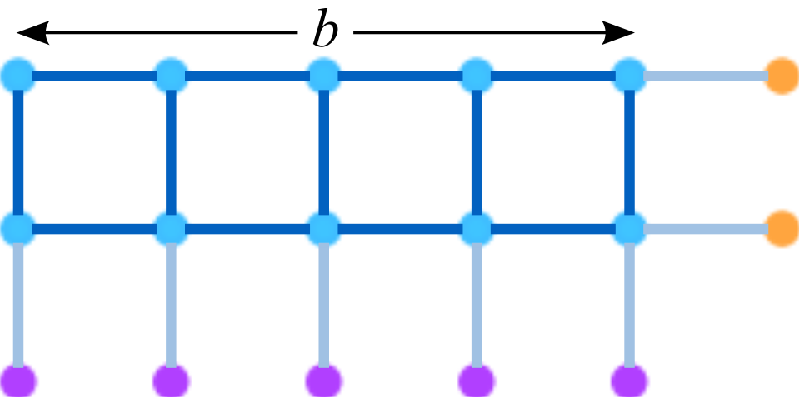}
        \label{fig2a}}
    \subfigure[]
       {\includegraphics[width=7.8cm]{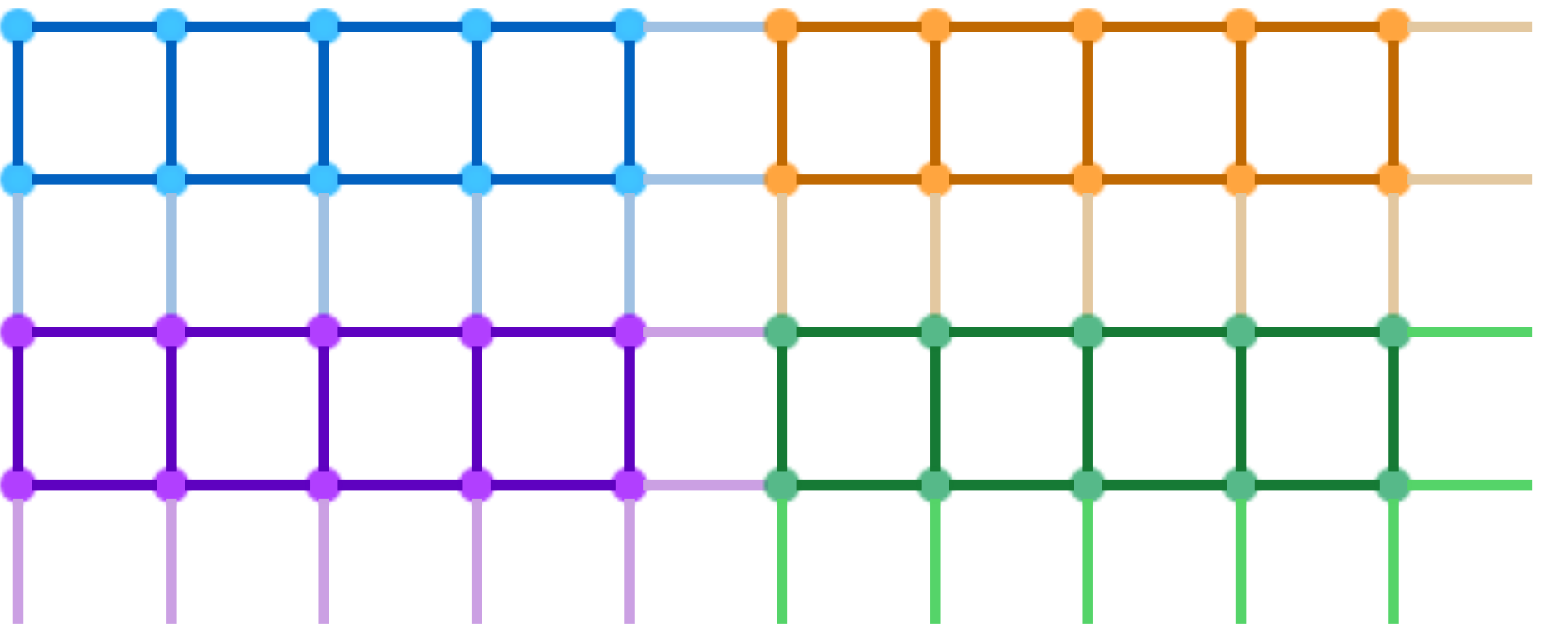}
        \label{fig2b}}
   \caption{(a) a Lego brick of size $b=5$, consisting of a core of 
	$2b=10$ qubits (blue) and $(b+2) = 7$ connections to qubits
	in neighbouring bricks,
	(b) 4 bricks joined together,
	showing how qubits are shared between bricks and thus entangled
	by more than one bus (at different times) during cluster construction
	(color online).}
   \label{fig2}
\end{figure}

We now consider the case where building the entire cluster with one
bus would take us beyond the threshold value of the error as given by
equation (\ref{eq:dephasing}). 
In such a situation we would need to use multiple buses, each one 
creating a smaller part of the cluster.
Since there are always at least two qubits entangled with the
bus for the path of width two, changing buses requires two qubit
disconnects and reconnects, a total of four extra operations per extra bus.
We term these sections of the cluster generated by one bus ``Lego bricks''
(figure \ref{fig2}).
If these Lego bricks have length $b$, see figure \ref{fig2a},
an $m\times n$ cluster will
contain $mn/2b$ of them.  The number of extra operations is thus
$4(mn/2b -1)$, giving a minimum number of operations
to create the cluster using multiple buses of 
\begin{equation}
N_{\rm{min}}(b) = (3 + 2/b)mn - 2(m+n).
\label{eq:bbound}
\end{equation}
Figure \ref{fig2b} shows how the bricks fit together in the
cluster, with shared qubits being reactivated by different buses
during the construction.
The bus thus entangles a total of $3b+2$ qubits to construct a brick that
adds $2b$ qubits to the cluster.
Where whole bricks fit neatly into the cluster (as shown),
we achieve the bound in equation (\ref{eq:bbound}).
Each brick needs $6b + 4$ bus operations to produce it
(two operations per qubit), so multiplying
by the number of bricks $mn/2b$, and subtracting the $2(m+n)$ operations
not required for the sides not connected to further bricks, 
the total number of operations $N_{\rm{min}}(b)$ is obtained.

\section{Dynamic generation}\label{sec:dynamic}

For dynamic generation of our cluster, we need to produce a
strip a few qubits wide with the measurements that perform the
computation applied just behind the construction process.
When a whole number of bricks fit across the cluster,
it can be dynamically generated without any loss of efficiency.
When bus changing operations don't happen conveniently at the edge
of the cluster, we will need to turn a corner within
a brick.  These U-shaped turns will cost at most two
extra operations per turn \cite{brown11b}.

\begin{figure}[t]
       \includegraphics[width=8.5cm]{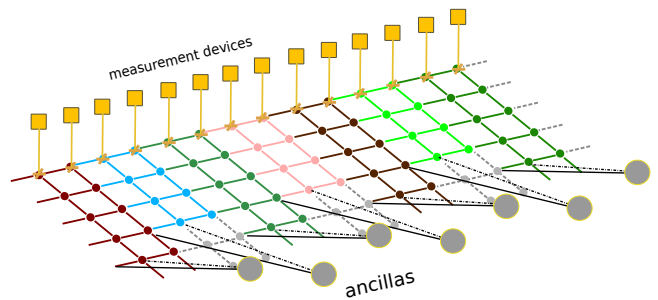}
    \caption{Dynamic generation using multiple ancillas (color online).}
   \label{fig3}
\end{figure}

Lego bricks also facilitate optimally efficient dynamic generation
where multiple buses are used in parallel to produce
a fully-scalable cluster-state scheme.  We orient our bricks
along the growth direction, see figure \ref{fig3},
producing parallel connected paths of width two.
To avoid the buses entangling to the
same qubit at the same time, alternate buses must be started six
operations apart.
Since we want a wide enough strip to allow room for the measurements
to follow behind the cluster construction, 
we can also use twice as many buses (one per qubit row).
This is less efficient in operations per bus, but generates a
wider strip in the same time frame \cite{brown11b}.
The optimal choice will depend on the decoherence rates and
the cost of extra buses for the particular system.

\section{Optimal brick size}\label{sec:threshold}

The system's error threshold $\varepsilon$ will determine the size of our Lego bricks.
A brick has $3b+2$ qubits, each
operated on twice for a total of $6b + 4$ bus operations, and
$4b$ edges (\textsc{cphase} gates); using equation 
(\ref{eq:dephasing}) we require
\begin{equation}
\frac{1}{2}\left(1 - \exp[-(6b+4) \gamma \tau - 16b \eta \beta^2]\right) \le \varepsilon.
\label{eq:etabricks}
\end{equation}
For a given set of experimental parameters $\gamma$, $\tau$ and $\eta$, and
desired dephasing limit $\varepsilon$, this determines $b$.

Let us now compare our scheme to the capabilities of one without bus 
reuse. If we use one bus per \textsc{cphase} gate to generate a brick, 
equation (\ref{eq:dephasing}) gives
\begin{equation}
\frac{1}{2}\left(1 - \exp[-16b \gamma \tau - 4 \eta \beta^2]\right) \le \varepsilon.
\label{eq:etasingle}
\end{equation}
Comparing equations (\ref{eq:etabricks}) and (\ref{eq:etasingle}),
we find our Lego scheme produces less qubit dephasing than using
one bus per \textsc{cphase} gate provided $\eta \beta^2 \lesssim \gamma \tau/2$.
For example, if $\gamma \tau = 5\times 10^{-4} $ and $\eta = 10^{-4} $,
then for an error threshold of $\varepsilon=10^{-2}$, the
bus-per-gate method could generate only 8 \textsc{cphase} gates between 8 qubits
($b=2$) before reaching the threshold, while the Lego method would be 
able to connect at least 17 qubits with 20 \textsc{cphase} gates ($b=5$)
before the same dephasing occurred.
For the case using multiple buses in parallel, this would give
a coherent strip of cluster four qubits wide, just enough to apply
the measurements behind the construction, as shown in figure \ref{fig3}.

\section{Conclusions}\label{sec:conc}

We have described the optimally efficient method for 
generating cluster states in ancilla-based computation, based on 
dividing the cluster into ``Lego bricks'', each of which is constructed 
with a single, reused, ancilla.  We have shown how, in the specific case 
of the qubus system, the reduction in ancilla operations can offset the 
increased noise due to bus reuse, allowing approximately twice the
number of qubits to be connected into a cluster state compared to
single-bus use.
Compared with $8mn - 4(m+n)$ bus operations with no bus reuse, 
for large clusters, the Lego scheme uses fewer than half for $b>2$, 
$O(3mn)$ compared to $O(8mn)$.  Even for $b=1$, the reduction is to 
$5mn - 2(m+n)$, equivalent to the method in \cite{Louis2007} when
limited to five qubits per bus.
This will therefore be the method of choice for any 
deterministic ancilla-based cluster generation that allows bus reuse
(see \cite{jens} for optimal probabilistic schemes).
This form of bus reuse can provide savings in many other
contexts, including the quantum Fourier transform \cite{brown11a}.

While the exact error model will vary with the underlying physical 
system, our analysis can be generalized to all ancilla-based cluster
generation schemes.
Our results are directly applicable to bus-based experimental 
production of cluster states, enabling the same resources to produce
dynamically generated cluster states of twice the size compared
to single-gate bus use.  For multibus dynamic schemes, this means
fully scalable operation can be achieved with half the coherence
time compared to single-gate buses.  In practical terms, this
needs as few as 20 \textsc{cphase} gates per bus, independent of cluster size.

\begin{acknowledgments}
\textit{We thank Aram Harrow, Ashley Stephens and Simon Devitt for helpful discussions.
CH was supported by EU project QAP and the Bristol Centre for Nanoscience and Quantum Information.
KLB is supported by a UK Engineering and Physical Sciences Research Council
industrial CASE studentship from Hewlett-Packard,
VMK is supported by a UK Royal Society University Research Fellowship and
WJM acknowledges partial support from the EU project HIP and MEXT in Japan.}
\end{acknowledgments}

\bibliography{../bibs/cluster_bibtex,../bibs/KLBbibliography,../bibs/KLBbibextra}

\begin{thebibliography}{31}%
\makeatletter
\providecommand \@ifxundefined [1]{%
 \@ifx{#1\undefined}
}%
\providecommand \@ifnum [1]{%
 \ifnum #1\expandafter \@firstoftwo
 \else \expandafter \@secondoftwo
 \fi
}%
\providecommand \@ifx [1]{%
 \ifx #1\expandafter \@firstoftwo
 \else \expandafter \@secondoftwo
 \fi
}%
\providecommand \natexlab [1]{#1}%
\providecommand \enquote  [1]{``#1''}%
\providecommand \bibnamefont  [1]{#1}%
\providecommand \bibfnamefont [1]{#1}%
\providecommand \citenamefont [1]{#1}%
\providecommand \href@noop [0]{\@secondoftwo}%
\providecommand \href [0]{\begingroup \@sanitize@url \@href}%
\providecommand \@href[1]{\@@startlink{#1}\@@href}%
\providecommand \@@href[1]{\endgroup#1\@@endlink}%
\providecommand \@sanitize@url [0]{\catcode `\\12\catcode `\$12\catcode
  `\&12\catcode `\#12\catcode `\^12\catcode `\_12\catcode `\%12\relax}%
\providecommand \@@startlink[1]{}%
\providecommand \@@endlink[0]{}%
\providecommand \url  [0]{\begingroup\@sanitize@url \@url }%
\providecommand \@url [1]{\endgroup\@href {#1}{\urlprefix }}%
\providecommand \urlprefix  [0]{URL }%
\providecommand \Eprint [0]{\href }%
\@ifxundefined \urlstyle {%
  \providecommand \doi  [0]{\begingroup \@sanitize@url \@doi}%
  \providecommand \@doi [1]{\endgroup \@@startlink {\doibase
  #1}doi:\discretionary {}{}{}#1\@@endlink }%
}{%
  \providecommand \doi  [0]{doi:\discretionary{}{}{}\begingroup
  \urlstyle{rm}\Url }%
}%
\providecommand \doibase [0]{http://dx.doi.org/}%
\providecommand \Doi [0]{\begingroup \@sanitize@url \@Doi }%
\providecommand \@Doi  [1]{\endgroup\@@startlink{\doibase#1}\@@Doi}%
\providecommand \@@Doi [1]{#1\@@endlink}%
\providecommand \selectlanguage [0]{\@gobble}%
\providecommand \bibinfo  [0]{\@secondoftwo}%
\providecommand \bibfield  [0]{\@secondoftwo}%
\providecommand \translation [1]{[#1]}%
\providecommand \BibitemOpen [0]{}%
\providecommand \bibitemStop [0]{}%
\providecommand \bibitemNoStop [0]{.\EOS\space}%
\providecommand \EOS [0]{\spacefactor3000\relax}%
\providecommand \BibitemShut  [1]{\csname bibitem#1\endcsname}%
\bibitem [{\citenamefont {Spiller}\ \emph {et~al.}(2006)\citenamefont
  {Spiller}, \citenamefont {Nemoto}, \citenamefont {Braunstein}, \citenamefont
  {Munro}, \citenamefont {van Loock},\ and\ \citenamefont {Milburn}}]{qubus06}%
  \BibitemOpen
  \bibfield  {author} {\bibinfo {author} {\bibfnamefont {T.~P.}\ \bibnamefont
  {Spiller}}, \bibinfo {author} {\bibfnamefont {K.}~\bibnamefont {Nemoto}},
  \bibinfo {author} {\bibfnamefont {S.~L.}\ \bibnamefont {Braunstein}},
  \bibinfo {author} {\bibfnamefont {W.~J.}\ \bibnamefont {Munro}}, \bibinfo
  {author} {\bibfnamefont {P.}~\bibnamefont {van Loock}}, \ and\ \bibinfo
  {author} {\bibfnamefont {G.~J.}\ \bibnamefont {Milburn}},\ }\Doi
  {10.1088/1367-2630/8/2/030} {\bibfield  {journal} {\bibinfo  {journal} {New
  J. Phys.},\ }\textbf {\bibinfo {volume} {8}},\ \bibinfo {pages} {30}
  (\bibinfo {year} {2006})}\BibitemShut {NoStop}%
\bibitem [{\citenamefont {van Loock}\ \emph {et~al.}(2008)\citenamefont {van
  Loock}, \citenamefont {Munro}, \citenamefont {Nemoto}, \citenamefont
  {Spiller}, \citenamefont {Ladd}, \citenamefont {Braunstein},\ and\
  \citenamefont {Milburn}}]{loockhybrid}%
  \BibitemOpen
  \bibfield  {author} {\bibinfo {author} {\bibfnamefont {P.}~\bibnamefont {van
  Loock}}, \bibinfo {author} {\bibfnamefont {W.~J.}\ \bibnamefont {Munro}},
  \bibinfo {author} {\bibfnamefont {K.}~\bibnamefont {Nemoto}}, \bibinfo
  {author} {\bibfnamefont {T.~P.}\ \bibnamefont {Spiller}}, \bibinfo {author}
  {\bibfnamefont {T.~D.}\ \bibnamefont {Ladd}}, \bibinfo {author}
  {\bibfnamefont {S.~L.}\ \bibnamefont {Braunstein}}, \ and\ \bibinfo {author}
  {\bibfnamefont {G.~J.}\ \bibnamefont {Milburn}},\ }\href@noop {} {\bibfield
  {journal} {\bibinfo  {journal} {Phys. Rev. A},\ }\textbf {\bibinfo {volume}
  {78}},\ \bibinfo {pages} {022303} (\bibinfo {year} {2008})}\BibitemShut
  {NoStop}%
\bibitem [{\citenamefont {Milburn}\ \emph {et~al.}(2000)\citenamefont
  {Milburn}, \citenamefont {Schneider},\ and\ \citenamefont
  {James}}]{milburn2000}%
  \BibitemOpen
  \bibfield  {author} {\bibinfo {author} {\bibfnamefont {G.~J.}\ \bibnamefont
  {Milburn}}, \bibinfo {author} {\bibfnamefont {S.}~\bibnamefont {Schneider}},
  \ and\ \bibinfo {author} {\bibfnamefont {D.~F.~V.}\ \bibnamefont {James}},\
  }\href@noop {} {\bibfield  {journal} {\bibinfo  {journal} {Fortschritte der
  Physik},\ }\textbf {\bibinfo {volume} {48}},\ \bibinfo {pages} {9} (\bibinfo
  {year} {2000})}\BibitemShut {NoStop}%
\bibitem [{\citenamefont {Anders}\ \emph {et~al.}(2010)\citenamefont {Anders},
  \citenamefont {Oi}, \citenamefont {Kashefi}, \citenamefont {Browne},\ and\
  \citenamefont {Andersson}}]{elham1}%
  \BibitemOpen
  \bibfield  {author} {\bibinfo {author} {\bibfnamefont {J.}~\bibnamefont
  {Anders}}, \bibinfo {author} {\bibfnamefont {D.~K.~L.}\ \bibnamefont {Oi}},
  \bibinfo {author} {\bibfnamefont {E.}~\bibnamefont {Kashefi}}, \bibinfo
  {author} {\bibfnamefont {D.~E.}\ \bibnamefont {Browne}}, \ and\ \bibinfo
  {author} {\bibfnamefont {E.}~\bibnamefont {Andersson}},\ }\href@noop {}
  {\bibfield  {journal} {\bibinfo  {journal} {Phys.~Rev.~A},\ }\textbf
  {\bibinfo {volume} {82}},\ \bibinfo {pages} {020301} (\bibinfo {year}
  {2010})}\BibitemShut {NoStop}%
\bibitem [{\citenamefont {Devitt}\ \emph {et~al.}(2007)\citenamefont {Devitt},
  \citenamefont {Greentree}, \citenamefont {Ionicioiu}, \citenamefont
  {O'Brien}, \citenamefont {Munro},\ and\ \citenamefont
  {Hollenberg}}]{Devitt2007}%
  \BibitemOpen
  \bibfield  {author} {\bibinfo {author} {\bibfnamefont {S.~J.}\ \bibnamefont
  {Devitt}}, \bibinfo {author} {\bibfnamefont {A.~D.}\ \bibnamefont
  {Greentree}}, \bibinfo {author} {\bibfnamefont {R.}~\bibnamefont
  {Ionicioiu}}, \bibinfo {author} {\bibfnamefont {J.~L.}\ \bibnamefont
  {O'Brien}}, \bibinfo {author} {\bibfnamefont {W.~J.}\ \bibnamefont {Munro}},
  \ and\ \bibinfo {author} {\bibfnamefont {L.~C.~L.}\ \bibnamefont
  {Hollenberg}},\ }\Doi {10.1103/PhysRevA.76.052312} {\bibfield  {journal}
  {\bibinfo  {journal} {Phys. Rev. A},\ }\textbf {\bibinfo {volume} {76}},\
  \bibinfo {pages} {052312} (\bibinfo {year} {2007})}\BibitemShut {NoStop}%
\bibitem [{\citenamefont {Ionicioiu}\ \emph {et~al.}(2009)\citenamefont
  {Ionicioiu}, \citenamefont {Spiller},\ and\ \citenamefont
  {Munro}}]{raducatalysis}%
  \BibitemOpen
  \bibfield  {author} {\bibinfo {author} {\bibfnamefont {R.}~\bibnamefont
  {Ionicioiu}}, \bibinfo {author} {\bibfnamefont {T.~P.}\ \bibnamefont
  {Spiller}}, \ and\ \bibinfo {author} {\bibfnamefont {W.~J.}\ \bibnamefont
  {Munro}},\ }\href@noop {} {\bibfield  {journal} {\bibinfo  {journal} {Phys.
  Rev. A},\ }\textbf {\bibinfo {volume} {80}},\ \bibinfo {pages} {012312}
  (\bibinfo {year} {2009})}\BibitemShut {NoStop}%
\bibitem [{\citenamefont {Devitt}\ \emph {et~al.}(2009)\citenamefont {Devitt},
  \citenamefont {Fowler}, \citenamefont {Stephens}, \citenamefont {Greentree},
  \citenamefont {Hollenberg}, \citenamefont {Munro},\ and\ \citenamefont
  {Nemoto}}]{simon2}%
  \BibitemOpen
  \bibfield  {author} {\bibinfo {author} {\bibfnamefont {S.~J.}\ \bibnamefont
  {Devitt}}, \bibinfo {author} {\bibfnamefont {A.~G.}\ \bibnamefont {Fowler}},
  \bibinfo {author} {\bibfnamefont {A.~M.}\ \bibnamefont {Stephens}}, \bibinfo
  {author} {\bibfnamefont {A.~D.}\ \bibnamefont {Greentree}}, \bibinfo {author}
  {\bibfnamefont {L.~C.}\ \bibnamefont {Hollenberg}}, \bibinfo {author}
  {\bibfnamefont {W.~J.}\ \bibnamefont {Munro}}, \ and\ \bibinfo {author}
  {\bibfnamefont {K.}~\bibnamefont {Nemoto}},\ }\Doi
  {10.1088/1367-2630/11/8/083032} {\bibfield  {journal} {\bibinfo  {journal}
  {New J. Phys.},\ }\textbf {\bibinfo {volume} {11}},\ \bibinfo {pages}
  {083032} (\bibinfo {year} {2009})}\BibitemShut {NoStop}%
\bibitem [{\citenamefont {Briegel}\ \emph {et~al.}(2009)\citenamefont
  {Briegel}, \citenamefont {Browne}, \citenamefont {Dür}, \citenamefont
  {Raussendorf},\ and\ \citenamefont {den Nest}}]{mbqc2}%
  \BibitemOpen
  \bibfield  {author} {\bibinfo {author} {\bibfnamefont {H.~J.}\ \bibnamefont
  {Briegel}}, \bibinfo {author} {\bibfnamefont {D.~E.}\ \bibnamefont {Browne}},
  \bibinfo {author} {\bibfnamefont {W.}~\bibnamefont {Dür}}, \bibinfo {author}
  {\bibfnamefont {R.}~\bibnamefont {Raussendorf}}, \ and\ \bibinfo {author}
  {\bibfnamefont {M.~V.}\ \bibnamefont {den Nest}},\ }\href@noop {} {\bibfield
  {journal} {\bibinfo  {journal} {Nature Physics},\ }\textbf {\bibinfo {volume}
  {5}},\ \bibinfo {pages} {19} (\bibinfo {year} {2009})}\BibitemShut {NoStop}%
\bibitem [{\citenamefont {Kitaev}(2003)}]{kitaev}%
  \BibitemOpen
  \bibfield  {author} {\bibinfo {author} {\bibfnamefont {A.~Y.}\ \bibnamefont
  {Kitaev}},\ }\href@noop {} {\bibfield  {journal} {\bibinfo  {journal} {Annals
  of Physics},\ }\textbf {\bibinfo {volume} {303}},\ \bibinfo {pages} {2}
  (\bibinfo {year} {2003})}\BibitemShut {NoStop}%
\bibitem [{\citenamefont {Raussendorf}\ and\ \citenamefont
  {Harrington}(2007)}]{surfacecode}%
  \BibitemOpen
  \bibfield  {author} {\bibinfo {author} {\bibfnamefont {R.}~\bibnamefont
  {Raussendorf}}\ and\ \bibinfo {author} {\bibfnamefont {J.}~\bibnamefont
  {Harrington}},\ }\href@noop {} {\bibfield  {journal} {\bibinfo  {journal}
  {Phys. Rev. Lett.},\ }\textbf {\bibinfo {volume} {98}},\ \bibinfo {pages}
  {190504} (\bibinfo {year} {2007})}\BibitemShut {NoStop}%
\bibitem [{\citenamefont {Devitt}\ \emph {et~al.}(2010)\citenamefont {Devitt},
  \citenamefont {Fowler}, \citenamefont {Tilma}, \citenamefont {Munro},\ and\
  \citenamefont {Nemoto}}]{devitt10a}%
  \BibitemOpen
  \bibfield  {author} {\bibinfo {author} {\bibfnamefont {S.~J.}\ \bibnamefont
  {Devitt}}, \bibinfo {author} {\bibfnamefont {A.~G.}\ \bibnamefont {Fowler}},
  \bibinfo {author} {\bibfnamefont {T.}~\bibnamefont {Tilma}}, \bibinfo
  {author} {\bibfnamefont {W.~J.}\ \bibnamefont {Munro}}, \ and\ \bibinfo
  {author} {\bibfnamefont {K.}~\bibnamefont {Nemoto}},\ }\href@noop {}
  {\bibfield  {journal} {\bibinfo  {journal} {Int J Quantum Inf},\ }\textbf
  {\bibinfo {volume} {8}},\ \bibinfo {pages} {121} (\bibinfo {year}
  {2010})}\BibitemShut {NoStop}%
\bibitem [{\citenamefont {Stephens}\ \emph {et~al.}(2008)\citenamefont
  {Stephens}, \citenamefont {Evans}, \citenamefont {Devitt}, \citenamefont
  {Greentree}, \citenamefont {Fowler}, \citenamefont {Munro}, \citenamefont
  {O'Brien}, \citenamefont {Nemoto},\ and\ \citenamefont
  {Hollenberg}}]{stephens08a}%
  \BibitemOpen
  \bibfield  {author} {\bibinfo {author} {\bibfnamefont {A.~M.}\ \bibnamefont
  {Stephens}}, \bibinfo {author} {\bibfnamefont {Z.~W.~E.}\ \bibnamefont
  {Evans}}, \bibinfo {author} {\bibfnamefont {S.~J.}\ \bibnamefont {Devitt}},
  \bibinfo {author} {\bibfnamefont {A.~D.}\ \bibnamefont {Greentree}}, \bibinfo
  {author} {\bibfnamefont {A.~G.}\ \bibnamefont {Fowler}}, \bibinfo {author}
  {\bibfnamefont {W.~J.}\ \bibnamefont {Munro}}, \bibinfo {author}
  {\bibfnamefont {J.~L.}\ \bibnamefont {O'Brien}}, \bibinfo {author}
  {\bibfnamefont {K.}~\bibnamefont {Nemoto}}, \ and\ \bibinfo {author}
  {\bibfnamefont {L.~C.~L.}\ \bibnamefont {Hollenberg}},\ }\href@noop {}
  {\bibfield  {journal} {\bibinfo  {journal} {Phys. Rev. A},\ }\textbf
  {\bibinfo {volume} {78}},\ \bibinfo {pages} {032318} (\bibinfo {year}
  {2008})}\BibitemShut {NoStop}%
\bibitem [{\citenamefont {Yoran}\ and\ \citenamefont
  {Reznik}(2003)}]{nadevreznik}%
  \BibitemOpen
  \bibfield  {author} {\bibinfo {author} {\bibfnamefont {N.}~\bibnamefont
  {Yoran}}\ and\ \bibinfo {author} {\bibfnamefont {B.}~\bibnamefont {Reznik}},\
  }\href@noop {} {\bibfield  {journal} {\bibinfo  {journal} {Phys. Rev.
  Lett.},\ }\textbf {\bibinfo {volume} {91}},\ \bibinfo {pages} {037903}
  (\bibinfo {year} {2003})}\BibitemShut {NoStop}%
\bibitem [{\citenamefont {Nielsen}(2004)}]{Nielsen2004}%
  \BibitemOpen
  \bibfield  {author} {\bibinfo {author} {\bibfnamefont {M.~A.}\ \bibnamefont
  {Nielsen}},\ }\Doi {10.1103/PhysRevLett.93.040503} {\bibfield  {journal}
  {\bibinfo  {journal} {Phys. Rev. Lett.},\ }\textbf {\bibinfo {volume} {93}},\
  \bibinfo {pages} {040503} (\bibinfo {year} {2004})}\BibitemShut {NoStop}%
\bibitem [{\citenamefont {Knill}\ \emph {et~al.}(2001)\citenamefont {Knill},
  \citenamefont {Laflamme},\ and\ \citenamefont {Milburn}}]{Knill2001}%
  \BibitemOpen
  \bibfield  {author} {\bibinfo {author} {\bibfnamefont {E.}~\bibnamefont
  {Knill}}, \bibinfo {author} {\bibfnamefont {R.}~\bibnamefont {Laflamme}}, \
  and\ \bibinfo {author} {\bibfnamefont {G.~J.}\ \bibnamefont {Milburn}},\
  }\Doi {10.1038/35051009} {\bibfield  {journal} {\bibinfo  {journal}
  {Nature},\ }\textbf {\bibinfo {volume} {409}},\ \bibinfo {pages} {46}
  (\bibinfo {year} {2001})}\BibitemShut {NoStop}%
\bibitem [{\citenamefont {Browne}\ and\ \citenamefont
  {Rudolph}(2005)}]{Browne2005}%
  \BibitemOpen
  \bibfield  {author} {\bibinfo {author} {\bibfnamefont {D.~E.}\ \bibnamefont
  {Browne}}\ and\ \bibinfo {author} {\bibfnamefont {T.}~\bibnamefont
  {Rudolph}},\ }\Doi {10.1103/PhysRevLett.95.010501} {\bibfield  {journal}
  {\bibinfo  {journal} {Phys.~Rev.~Lett.},\ }\textbf {\bibinfo {volume} {95}},\
  \bibinfo {pages} {010501} (\bibinfo {year} {2005})}\BibitemShut {NoStop}%
\bibitem [{\citenamefont {Duan}\ and\ \citenamefont
  {Raussendorf}(2005)}]{Duan2005}%
  \BibitemOpen
  \bibfield  {author} {\bibinfo {author} {\bibfnamefont {L.-M.}\ \bibnamefont
  {Duan}}\ and\ \bibinfo {author} {\bibfnamefont {R.}~\bibnamefont
  {Raussendorf}},\ }\Doi {10.1103/PhysRevLett.95.080503} {\bibfield  {journal}
  {\bibinfo  {journal} {Phys. Rev. Lett.},\ }\textbf {\bibinfo {volume} {95}},\
  \bibinfo {pages} {080503} (\bibinfo {year} {2005})}\BibitemShut {NoStop}%
\bibitem [{\citenamefont {Gross}\ \emph {et~al.}(2006)\citenamefont {Gross},
  \citenamefont {Kieling},\ and\ \citenamefont {Eisert}}]{jens}%
  \BibitemOpen
  \bibfield  {author} {\bibinfo {author} {\bibfnamefont {D.}~\bibnamefont
  {Gross}}, \bibinfo {author} {\bibfnamefont {K.}~\bibnamefont {Kieling}}, \
  and\ \bibinfo {author} {\bibfnamefont {J.}~\bibnamefont {Eisert}},\
  }\href@noop {} {\bibfield  {journal} {\bibinfo  {journal} {Phys. Rev. A},\
  }\textbf {\bibinfo {volume} {74}},\ \bibinfo {pages} {042343} (\bibinfo
  {year} {2006})}\BibitemShut {NoStop}%
\bibitem [{\citenamefont {Louis}\ \emph
  {et~al.}(2007){\natexlab{a}}\citenamefont {Louis}, \citenamefont {Nemoto},
  \citenamefont {Munro},\ and\ \citenamefont {Spiller}}]{Louis2007}%
  \BibitemOpen
  \bibfield  {author} {\bibinfo {author} {\bibfnamefont {S.~G.~R.}\
  \bibnamefont {Louis}}, \bibinfo {author} {\bibfnamefont {K.}~\bibnamefont
  {Nemoto}}, \bibinfo {author} {\bibfnamefont {W.~J.}\ \bibnamefont {Munro}}, \
  and\ \bibinfo {author} {\bibfnamefont {T.~P.}\ \bibnamefont {Spiller}},\
  }\Doi {10.1088/1367-2630/9/6/193} {\bibfield  {journal} {\bibinfo  {journal}
  {New Journal of Physics},\ }\textbf {\bibinfo {volume} {9}},\ \bibinfo
  {pages} {193} (\bibinfo {year} {2007}{\natexlab{a}})}\BibitemShut {NoStop}%
\bibitem [{\citenamefont {Hutchinson}\ and\ \citenamefont
  {Milburn}(2004)}]{hutchinson04}%
  \BibitemOpen
  \bibfield  {author} {\bibinfo {author} {\bibfnamefont {G.~D.}\ \bibnamefont
  {Hutchinson}}\ and\ \bibinfo {author} {\bibfnamefont {G.~J.}\ \bibnamefont
  {Milburn}},\ }\href@noop {} {\bibfield  {journal} {\bibinfo  {journal}
  {J.~Mod.~Optics},\ }\textbf {\bibinfo {volume} {51}},\ \bibinfo {pages}
  {1211} (\bibinfo {year} {2004})}\BibitemShut {NoStop}%
\bibitem [{\citenamefont {Lin}\ and\ \citenamefont {He}(2010)}]{Lin2010}%
  \BibitemOpen
  \bibfield  {author} {\bibinfo {author} {\bibfnamefont {Q.}~\bibnamefont
  {Lin}}\ and\ \bibinfo {author} {\bibfnamefont {B.}~\bibnamefont {He}},\ }\Doi
  {10.1103/PhysRevA.82.022331} {\bibfield  {journal} {\bibinfo  {journal}
  {Phys. Rev. A},\ }\textbf {\bibinfo {volume} {82}},\ \bibinfo {pages}
  {022331} (\bibinfo {year} {2010})}\BibitemShut {NoStop}%
\bibitem [{\citenamefont {Ionicioiu}\ and\ \citenamefont
  {Munro}(2010)}]{Ionicioiu10a}%
  \BibitemOpen
  \bibfield  {author} {\bibinfo {author} {\bibfnamefont {R.}~\bibnamefont
  {Ionicioiu}}\ and\ \bibinfo {author} {\bibfnamefont {W.~J.}\ \bibnamefont
  {Munro}},\ }\href@noop {} {\bibfield  {journal} {\bibinfo  {journal}
  {International Journal of Quantum Information},\ }\textbf {\bibinfo {volume}
  {8}},\ \bibinfo {pages} {149} (\bibinfo {year} {2010})}\BibitemShut {NoStop}%
\bibitem [{\citenamefont {Wang}\ \emph
  {et~al.}(2010){\natexlab{a}}\citenamefont {Wang}, \citenamefont {Yang},\ and\
  \citenamefont {Nori}}]{Wang2010}%
  \BibitemOpen
  \bibfield  {author} {\bibinfo {author} {\bibfnamefont {H.}~\bibnamefont
  {Wang}}, \bibinfo {author} {\bibfnamefont {C.~P.}\ \bibnamefont {Yang}}, \
  and\ \bibinfo {author} {\bibfnamefont {F.}~\bibnamefont {Nori}},\ }\Doi
  {10.1103/PhysRevA.81.052332} {\bibfield  {journal} {\bibinfo  {journal}
  {Phys. Rev. A},\ }\textbf {\bibinfo {volume} {81}},\ \bibinfo {pages}
  {052332} (\bibinfo {year} {2010}{\natexlab{a}})}\BibitemShut {NoStop}%
\bibitem [{\citenamefont {Fowler}\ \emph {et~al.}(2009)\citenamefont {Fowler},
  \citenamefont {Stephens},\ and\ \citenamefont {Groszkowski}}]{austin1}%
  \BibitemOpen
  \bibfield  {author} {\bibinfo {author} {\bibfnamefont {A.~G.}\ \bibnamefont
  {Fowler}}, \bibinfo {author} {\bibfnamefont {A.~M.}\ \bibnamefont
  {Stephens}}, \ and\ \bibinfo {author} {\bibfnamefont {P.}~\bibnamefont
  {Groszkowski}},\ }\href@noop {} {\bibfield  {journal} {\bibinfo  {journal}
  {Phys. Rev. A},\ }\textbf {\bibinfo {volume} {80}},\ \bibinfo {pages}
  {052312} (\bibinfo {year} {2009})}\BibitemShut {NoStop}%
\bibitem [{\citenamefont {Wang}\ \emph
  {et~al.}(2010){\natexlab{b}}\citenamefont {Wang}, \citenamefont {Fowler},
  \citenamefont {Stephens},\ and\ \citenamefont {Hollenberg}}]{austin2}%
  \BibitemOpen
  \bibfield  {author} {\bibinfo {author} {\bibfnamefont {D.~S.}\ \bibnamefont
  {Wang}}, \bibinfo {author} {\bibfnamefont {A.~G.}\ \bibnamefont {Fowler}},
  \bibinfo {author} {\bibfnamefont {A.~M.}\ \bibnamefont {Stephens}}, \ and\
  \bibinfo {author} {\bibfnamefont {L.~C.~L.}\ \bibnamefont {Hollenberg}},\
  }\href@noop {} {\bibfield  {journal} {\bibinfo  {journal} {Quantum
  Information and Computation},\ }\textbf {\bibinfo {volume} {10}},\ \bibinfo
  {pages} {456} (\bibinfo {year} {2010}{\natexlab{b}})}\BibitemShut {NoStop}%
\bibitem [{\citenamefont {Louis}\ \emph {et~al.}(2008)\citenamefont {Louis},
  \citenamefont {Munro}, \citenamefont {Spiller},\ and\ \citenamefont
  {Nemoto}}]{qubus3}%
  \BibitemOpen
  \bibfield  {author} {\bibinfo {author} {\bibfnamefont {S.~G.~R.}\
  \bibnamefont {Louis}}, \bibinfo {author} {\bibfnamefont {W.~J.}\ \bibnamefont
  {Munro}}, \bibinfo {author} {\bibfnamefont {T.~P.}\ \bibnamefont {Spiller}},
  \ and\ \bibinfo {author} {\bibfnamefont {K.}~\bibnamefont {Nemoto}},\
  }\href@noop {} {\bibfield  {journal} {\bibinfo  {journal} {Phys. Rev A},\
  }\textbf {\bibinfo {volume} {78}},\ \bibinfo {pages} {022326} (\bibinfo
  {year} {2008})}\BibitemShut {NoStop}%
\bibitem [{\citenamefont {Munro}\ \emph {et~al.}(2005)\citenamefont {Munro},
  \citenamefont {Nemoto}, \citenamefont {Spiller}, \citenamefont {Barrett},
  \citenamefont {Kok},\ and\ \citenamefont {Beausoleil}}]{qubus2}%
  \BibitemOpen
  \bibfield  {author} {\bibinfo {author} {\bibfnamefont {W.~J.}\ \bibnamefont
  {Munro}}, \bibinfo {author} {\bibfnamefont {K.}~\bibnamefont {Nemoto}},
  \bibinfo {author} {\bibfnamefont {T.~P.}\ \bibnamefont {Spiller}}, \bibinfo
  {author} {\bibfnamefont {S.~D.}\ \bibnamefont {Barrett}}, \bibinfo {author}
  {\bibfnamefont {P.}~\bibnamefont {Kok}}, \ and\ \bibinfo {author}
  {\bibfnamefont {R.~G.}\ \bibnamefont {Beausoleil}},\ }\href@noop {}
  {\bibfield  {journal} {\bibinfo  {journal} {J. Opt. B: Quantum Semiclass.
  Opt.},\ }\textbf {\bibinfo {volume} {7}},\ \bibinfo {pages} {S135} (\bibinfo
  {year} {2005})}\BibitemShut {NoStop}%
\bibitem [{\citenamefont {Wang}\ and\ \citenamefont
  {Zanardi}(2002)}]{wangzanardi}%
  \BibitemOpen
  \bibfield  {author} {\bibinfo {author} {\bibfnamefont {X.}~\bibnamefont
  {Wang}}\ and\ \bibinfo {author} {\bibfnamefont {P.}~\bibnamefont {Zanardi}},\
  }\href@noop {} {\bibfield  {journal} {\bibinfo  {journal} {Phys.~Rev.~A},\
  }\textbf {\bibinfo {volume} {65}},\ \bibinfo {pages} {032327} (\bibinfo
  {year} {2002})}\BibitemShut {NoStop}%
\bibitem [{\citenamefont {Louis}\ \emph
  {et~al.}(2007){\natexlab{b}}\citenamefont {Louis}, \citenamefont {Nemoto},
  \citenamefont {Munro},\ and\ \citenamefont {Spiller}}]{louis2}%
  \BibitemOpen
  \bibfield  {author} {\bibinfo {author} {\bibfnamefont {S.~G.~R.}\
  \bibnamefont {Louis}}, \bibinfo {author} {\bibfnamefont {K.}~\bibnamefont
  {Nemoto}}, \bibinfo {author} {\bibfnamefont {W.~J.}\ \bibnamefont {Munro}}, \
  and\ \bibinfo {author} {\bibfnamefont {T.~P.}\ \bibnamefont {Spiller}},\
  }\href@noop {} {\bibfield  {journal} {\bibinfo  {journal} {Phys. Rev. A},\
  }\textbf {\bibinfo {volume} {75}},\ \bibinfo {pages} {042323} (\bibinfo
  {year} {2007}{\natexlab{b}})}\BibitemShut {NoStop}%
\bibitem [{\citenamefont {Brown}\ \emph {et~al.}(2011)\citenamefont {Brown},
  \citenamefont {Horsman}, \citenamefont {Munro},\ and\ \citenamefont
  {Kendon}}]{brown11b}%
  \BibitemOpen
  \bibfield  {author} {\bibinfo {author} {\bibfnamefont {K.~L.}\ \bibnamefont
  {Brown}}, \bibinfo {author} {\bibfnamefont {C.}~\bibnamefont {Horsman}},
  \bibinfo {author} {\bibfnamefont {W.~J.}\ \bibnamefont {Munro}}, \ and\
  \bibinfo {author} {\bibfnamefont {V.~M.}\ \bibnamefont {Kendon}},\
  }\href@noop {} {\enquote {\bibinfo {title} {{Lego} brick construction of
  cluster states.}}\ } (\bibinfo {year} {2011}),\ \bibinfo {note}
  {unpublished.}\BibitemShut {Stop}%
\bibitem [{\citenamefont {Brown}\ \emph {et~al.}(2010)\citenamefont {Brown},
  \citenamefont {De}, \citenamefont {Munro},\ and\ \citenamefont
  {Kendon}}]{brown11a}%
  \BibitemOpen
  \bibfield  {author} {\bibinfo {author} {\bibfnamefont {K.~L.}\ \bibnamefont
  {Brown}}, \bibinfo {author} {\bibfnamefont {S.}~\bibnamefont {De}}, \bibinfo
  {author} {\bibfnamefont {W.~J.}\ \bibnamefont {Munro}}, \ and\ \bibinfo
  {author} {\bibfnamefont {V.~M.}\ \bibnamefont {Kendon}},\ }\href@noop {}
  {\enquote {\bibinfo {title} {Ancilla-based quantum simulation},}\ } (\bibinfo
  {year} {2010}),\ \Eprint {http://arxiv.org/abs/arXiv:1011.2984v4[quant-ph]}
  {arXiv:1011.2984v4[quant-ph]} \BibitemShut {NoStop}%
\end{thebibliography}%

\end{document}